# Dual origin of defect magnetism in graphene and its reversible switching by molecular doping

R. R. Nair[1], I-Ling Tsai[1], M. Sepioni[1], O. Lehtinen[2], J. Keinonen[2], A. V. Krasheninnikov[2,3],

A. H. Castro Neto[4], M. I. Katsnelson[5], A. K. Geim[1], I. V. Grigorieva[1*]

[1]Manchester Centre for Mesoscience & Nanotechnology, Manchester M13 9PL, UK
[2]Department of Physics, P.O. Box 43, FI-00014 University of Helsinki, Finland
[3]Department of Applied Physics, P.O. Box 11100, Aalto University, FI-00076, Finland
[4]Graphene Research Centre and Department of Physics, National University of Singapore, 6 Science Dr. 2, Singapore, 117456
[5]Radboud University of Nijmegen, Institute for Molecules and Materials, 6525 AJ Nijmegen, The Netherlands

*correspondence should be addressed to I.V.G.:  irina.grigorieva@manchester.ac.uk

*A possibility to control magnetic properties by using electric fields is one of the most desirable characteristics for spintronics applications. Finding a suitable material remains an elusive goal, with only a few candidates found so far. Graphene is one of them and offers a hope due to its weak spin-orbit interaction, the ability to control electronic properties by the electric field effect and the possibility to introduce paramagnetic centres such as vacancies and adatoms. Here we show that adatoms' magnetism in graphene is itinerant and can be controlled by doping, so that magnetic moments can be switched on and off. The much-discussed vacancy magnetism is found to have a dual origin, with two approximately equal contributions: one coming from the same itinerant magnetism and the other due to broken bonds. Our work suggests that graphene's magnetism can be controlled by the field effect, similar to its transport and optical properties, and that spin diffusion length can be significantly enhanced above a certain carrier density.*

Electric–field control of magnetic properties has been subject of considerable attention, with a few materials and hybrid magnetic systems found recently where it is possible to control the magnetisation direction and/or the Curie temperature [1-3]. Electric field tunability is one of the fundamental properties of graphene and has been widely used to control its electronic, optical and other properties related to the electronic structure [4]. On the other hand, graphene is believed to be an ideal material for spintronics [5] due to the weak spin-orbit interaction and long spin relaxation lengths [6,7] and the possibility to introduce paramagnetic centres via controlled introduction of defects, as recent experiments [8,9] have shown that both vacancies and adatoms in graphene carry magnetic moments $\mu \approx \mu_B$. In principle, the latter is not exceptional because defects and impurities in crystals without any $d$ or $f$ elements may have unpaired electrons and, therefore, exhibit paramagnetism [10,11]. However, achieving ferromagnetic alignment between these unpaired spins has proven practically impossible [10]. In graphene the prospect of defect ferromagnetism is more realistic because, firstly, the presence of conduction electrons provides a medium for coupling between localised spins and, secondly, the defect-induced moments are believed to be due to the same π electrons that are responsible for electron transport [12-21]. Accordingly, it should be possible to use the electric-field effect to control not only the coupling between localised magnetic moments in graphene, but also the presence of the moments themselves which so far has not been possible in other materials.

In this contribution we have employed SQUID magnetometry and molecular doping to investigate graphene's paramagnetism (here SQUID stands for Superconducting Quantum Interference Device). We have found that magnetic moments due to vacancies and adatoms can be switched off by shifting the Fermi energy $E_F$ above ≈0.45eV and then switched back on by returning to the neutral state or lower $E_F$. This proves the itinerant nature of magnetic moments in graphene, that is, π magnetism. The localization radius $r_L$ for the itinerant spins is estimated as ≈2.0 nm. For the case of vacancies, we find them carrying two practically independent moments that are attributed to the π magnetism and an unpaired spin on broken bonds. Only half of vacancy magnetism can be switched off by doping. A profound implication of our findings is that π magnetism can be controlled by other means, notably by the electric field doping within realistically accessible carrier densities $n$ ≈$2 \times 10^{13}$cm$^{-2}$. Furthermore, it is likely that the unexpectedly quick spin relaxation found in graphene devices and attributed to magnetic defects [22] can be quenched by doping above the above threshold.

Our samples were graphene laminates that were shown [8] to be a well-characterized, clean reference system, providing enough material (typically, several mg of graphene) for SQUID magnetometry. Our magnetometer (Quantum Design MPMS XL7) allowed measurements at temperatures $T$ between 300 and 1.8 K in fields $H$ up to 7T. The samples were irradiated with 350 keV protons to achieve the desired density of single vacancies, $n_v$ (Methods). Irradiation resulted in a pronounced paramagnetic response arising from non-interacting moments [8]. Typical magnetisation curves are shown in Fig. 1. For all vacancy concentrations, such curves were accurately described by the Brillouin function with spin $S$=1/2. The measured magnetic moment $M$ was found to be proportional to $n_v$ as $M \propto N_S \cdot \mu_B \cdot S \propto n_v$, where $\mu_B$ is the Bohr magneton and $N_S$ the number of spins extracted from Brillouin function fits, in agreement with the earlier report [8].

To probe the evolution of $M$ as the valence(conduction) band gets progressively filled up with holes(electrons), we varied the carrier concentration by using molecular doping: nitric acid $HNO_3$ or $NO_2$ gas as hole dopants [23-25] and aniline as an electron dopant [26]. The properties of $HNO_3$ and $NO_2$ as acceptors were investigated both theoretically [24] and experimentally [23,25] and both were shown to be very effective, with each physisorbed molecule resulting in transfer of one electron [25] (see Methods for details). However, finding an effective donor turned out to be problematic. Aniline is reported as one of the best donors but the maximum doping it allowed was only $n \approx 5 \cdot 10^{12}$ cm$^{-2}$. Within the limited range of $n$ achievable for aniline, we have found no difference with respect to hole doping with HNO$_3$ and, therefore, focus below on the latter results.

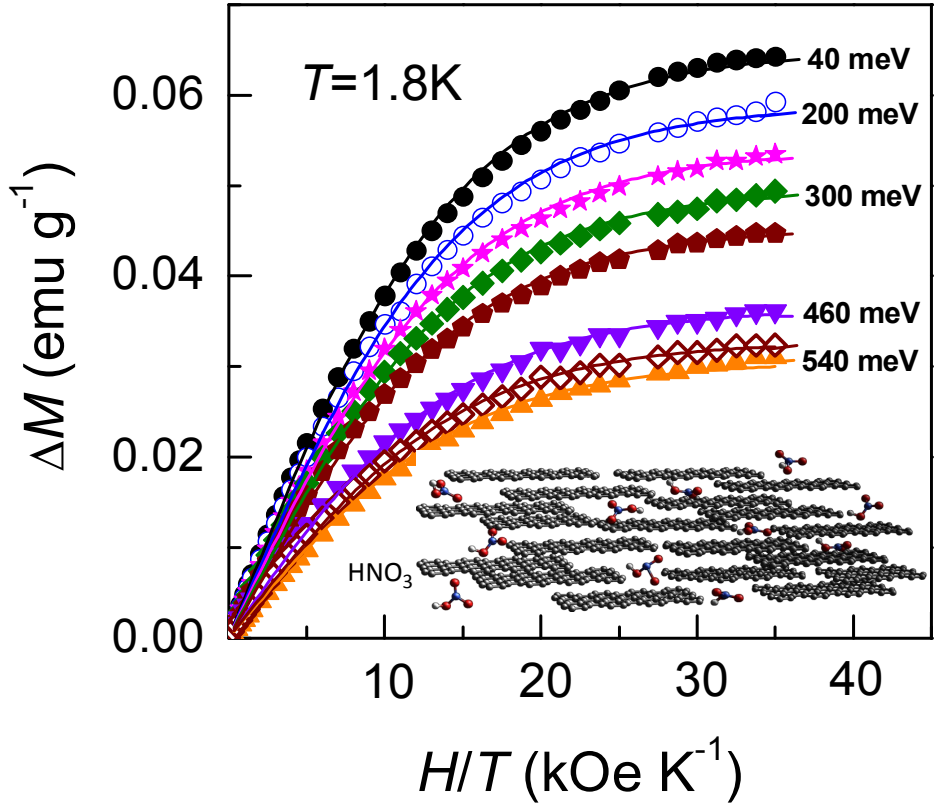

*Figure 1.* **Effect of hole concentration on vacancy paramagnetism.** Symbols show the measured magnetic moment $\Delta M$ as a function of magnetic field $H$ for different $n$. A linear diamagnetic background is subtracted. Solid curves are fits to the Brillouin function with $S = 1/2$. Labels show the corresponding $|E_F|$ in the valence band. Inset: Illustration of HNO$_3$ molecules physisorbed in graphene laminates.

Figure 1 shows the evolution of $M(H)$ curves for one of the irradiated samples as it is doped from a nearly neutral state to $E_F \approx 0.5$eV. It is clear that the doping has a strong effect on magnetisation: $M$ reduces to half its initial value, even though all $M(H)$ curves look qualitatively the same and are accurately described by the Brillouin function with $S=1/2$. This indicates a reduction in the number of magnetic moments for the same number of vacancies. It also shows that high $n$ do not lead to stronger interaction, as the vacancy magnetic moments remain non-interacting. This behaviour was found to be universal, even though $n_v$ varied for different samples by more than an order of magnitude, with the average separation between vacancies changing from $\approx 3$ to 10 nm. Fig. 2 shows the evolution of the number of detected spins $N_S$ in two graphene samples with different $n_v$ as a function of $E_F$. Initially, $N_S$ is gradually decreasing with increasing the hole concentration but at $E_F \approx 0.4$eV it falls sharply and seems to saturate to $N_S \approx N_S^0/2$ at the maximum doping. As soon as the dopants were removed from graphene by mild annealing, $M$ recovered its initial value, i.e. the effect is fully reversible (Fig. 2).

$E_F \approx 0.54$eV was the maximum doping that we were able to achieve with HNO$_3$, in agreement with the calculated energy dependence of the density of states for the NO$_2$ molecule adsorbed on graphene [24]. The data scatter between different samples and doping levels does not allow us to follow the transition in Fig. 2 in more detail and confirm the apparent saturation to half magnetization above $E_F \approx 0.45$eV. However, this behaviour was observed for many samples and repeatedly after annealing cycles. To gain further insight, Fig. 3 plots the reduction in magnetization signal at maximum doping, which was measured for several samples with different vacancy densities. It is clear that, independent of the sample and its $N_S^0$, the 50% reduction in $M$ is universal, as long as the Fermi level is shifted sufficiently away from the neutrality point.

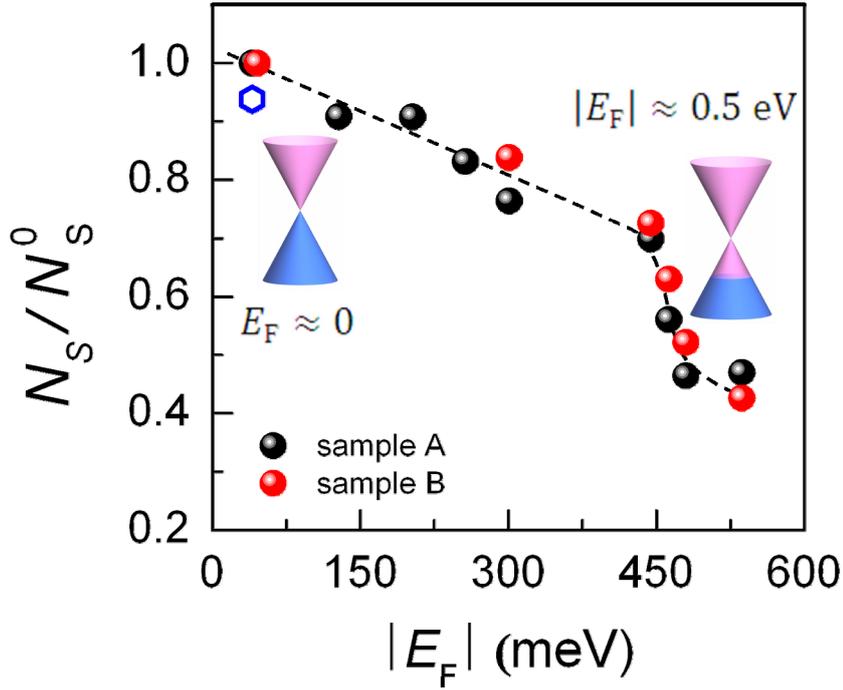

*Figure 2.* **Reversible control of vacancy magnetism.** Normalised density of spins ($S$=1/2) in two irradiated samples as a function of their doping.     is the spin concentration prior to doping [6], and     is extracted from the Brillouin function fit for the $M(H)$ dependence for each doping level. The dashed curve is a guide to the eye. Open symbol is for sample B after it was maximally doped and then annealed at ~100°C for 2h. The insets are schematic illustrations of the graphene spectrum in neutral and heavily doped states.

We attribute this finding to the presence of two types of magnetic moments associated with a vacancy and only one of them is being affected by such a shift in $E_F$. Indeed, let us recall the origin of defect-induced magnetism in graphene. It lies in the imbalance that the defects create between its two sublattices and the resulting appearance of localised π states with energy close to the Dirac point [12-21]. As long as these states are singly occupied, due to Coulomb repulsion, $U_C$, they give rise to magnetic moments μ=1μ$_B$. This simple picture is expected for adatoms, such as fluorine or hydrogen, whereas for vacancies the origin of magnetic moments is more complex. In addition to the above π moments, there are dangling σ bonds that can play some role, too [14,16,18-21]. For example, vacancies in insulating boron nitride are expected to be magnetic [27]. The relative contribution of σ states to vacancy magnetism is unknown and a matter of debate [18,20]. Due to high chemical reactivity of the dangling bonds, it is often assumed that they are saturated with, e.g., hydrogen and therefore do not contribute to magnetism (see e.g. refs. [18,21]). Then, the magnetic moment is purely due to itinerant (conduction) electrons. Another possible scenario is a vacancy reconstruction and the associated Jan-Teller distortion that results in hybridisation between the π and σ states [14,19-21]. Depending on the degree of hybridisation, the magnetic moment for a vacancy has been predicted to be between ~1.45 and 2 $\mu_B$ [14,16,19,21].

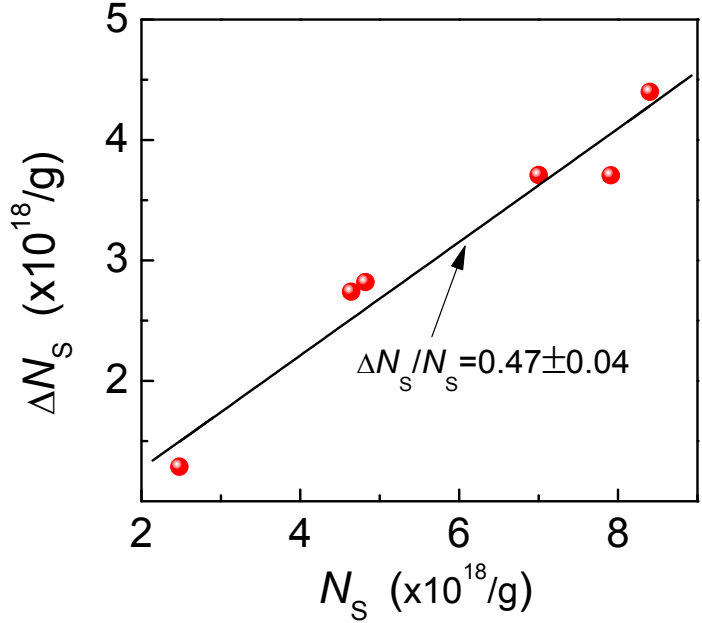

*Figure 3*. **Universal halving of the vacancy magnetic moment by doping.** Symbols show the change in the number of spins after maximum doping, $\Delta N_S = N_S^0 - N_S^f$ in samples with different $n_v$. $N_S^f$ is measured at $E_F \geq 0.5$ eV. Solid line is the best fit $\Delta N_S/N_S^0 = 0.47\pm0.04$.

Our experiment suggests an intermediate scenario in which the dangling bonds remain unsaturated and the hybridization does not play a significant role. In this case, each vacancy is expected to provide two independent contributions, which come from the singly-occupied localized π state and the unsaturated σ bond. At sufficiently large doping, such that $|E_F| \geq U_C$, the localised π state becomes doubly occupied and the corresponding magnetic moment disappears, whereas the unpaired σ electron remains unaffected by doping and provides the residual $M$. Since the two contributions appear to be nearly equal, hybridisation between the two electronic states should be weak. This is also in agreement with the measured $S = 1/2$. Note that it would require unrealistically high doping to occupy the σ state and fully suppress the vacancy magnetism.

To support the above interpretation, a neat experiment would be to compare the effect of doping on vacancy magnetism with that for adatoms, where there are no dangling bonds and only localised π states are expected [13-15]. However, it has proven difficult experimentally to introduce adatoms without their significant clustering. The latter locally opens a band gap, quenches nearest-neighbour magnetic moments, etc. so that the system can no longer be considered as graphene but becomes its chemical derivative [8]. We have circumvented this problem by exploiting the experimental fact that, if graphene is heated at ≥350°C, it results in the appearance of so-called resonant scatterers [28-31]. They reduce graphene's electronic quality and result in the D peak in Raman spectra. The scatterers are attributed to some organic residue that is always present on graphene and strongly ($sp^3$) bonds to it at high $T$ [28]. Resonant scatterers is another name for the localised π-states near the Dirac point in the context of electron transport. With these earlier studies in mind, we annealed several samples of pristine (non-irradiated) graphene laminates at 350°C in Ar-$H_2$ atmosphere. This induced notable paramagnetism with $N_S^0 \approx 3 \cdot 10^{18}$ g$^{-1}$. The observed $M(H)$ curves were practically identical to those of irradiated laminates and, again, are accurately described by the Brillouin function with $S=1/2$. The found $N_S^0$ corresponds to 2-3 $sp^3$ bonds induced by annealing per each graphene crystallite in the laminates.

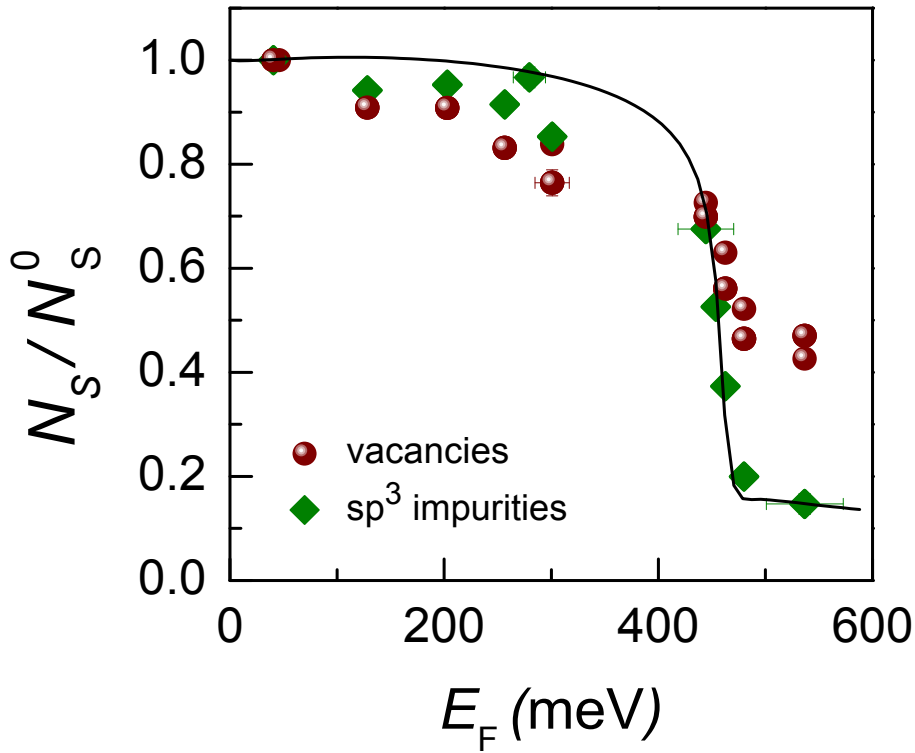

*Figure 4*. **Fully switchable π magnetism of adatoms.** Comparison of the effect of doping on magnetism associated with vacancies (data from Fig. 2) and with $sp^3$ defects. Solid curve is the theoretical dependence $N_S/N_S^0(E_F)$ calculated using equations (1) and (2), see text. Error bars indicate the accuracy of determination of $E_F$.

In contrast to vacancy magnetism, the exposure of these $sp^3$-defected samples to $HNO_3$ resulted in an almost complete disappearance of the paramagnetic signal above the same $E_F \approx 0.45$eV (Fig. 4), in agreement with expectations for the purely itinerant nature of $sp^3$ magnetism. Similar to the case of vacancies, the switching effect is also fully reversible: as soon as $HNO_3$ molecules were removed by mild annealing at ~100°C, the paramagnetic signal recovered its initial value $N_S^0$.

The magnetic moment due to $sp^3$ impurities and its variation with doping can be calculated analytically using the expression

$$N(E) = \frac{a^2 |E|}{\left[E\left(1 - a^2 \ln\left|\frac{E^2}{1-E^2}\right|\right) + E_d\right]^2 + \left(\pi a^2 E\right)^2} \quad (1)$$

which describes the density of states (DOS) in graphene due to resonant scatterers such as sp³ adsorbates (see Supplementary Note 2). Here, $E_d$ is the states' energy position and $a$ the hybridization parameter, both in units of the half width $W \approx 3t$ of graphene' energy band ($t$ is the nearest neighbor hopping parameter). For sp³ impurities relevant to our experiment (monovalent organic groups or hydrogen adatoms), $E_d$ =0.02 and $a$ =2/3 (that is, hybridization strength $\approx 2t$) [31]. The magnetic moment (number of spins) is given by the fraction of singly occupied impurity states, which is determined by the Hubbard energy $U$ and can be found by integrating (1) over the occupied states:

$$N_S(E_F) = \int_{E_F - U}^{E_F} dE\, N(E) \quad (2)$$

Using (2), we obtained the $N_S(E_F)$ shown in Fig. 4, which effectively has only one fitting parameter, $U$, because the dependence is found to be qualitatively insensitive to $E_d$ and $a$ over a wide range of the parameters (the best-fitting value of $U$ as in Fig. 4 is $U$ =0.5 eV). $N_S$ drops sharply at $E_F \approx U$ as expected, exhibiting a cusp-like dependence $1/\ln|E_F - U|$. Note that the cut-off does not lead to zero magnetization, which seems surprising but is in agreement with the experiment. The reason for a finite remaining $N_S$ on the theoretical curve is that the density of states deviates significantly from the standard Lorentzian peak. This arises due to the linear increase in the density of states of conduction electrons in graphene and a renormalization of the positions of impurity states because of hybridization. As a result, $N(E)$ decays as $1/|E|\ln^2|E|$ at large $E$, much slower than the standard $1/E^2$ dependence for the Lorentz distribution. This means that the magnetic states become totally doubly occupied only for $E_F \gg U$ and, therefore, $N_S$ exhibits a long tail at higher $E_F$ such as in Fig. 4.

Finally, let us add several comments. First, the found value of the Coulomb energy $U$=0.5 eV allows us to estimate the spatial localisation $r_L$ of the π states as $r_L \approx 2.0$ nm (using the known effective Coulomb interactions in graphene [32]), i.e. the π state is localised over several benzene rings, in agreement with theory [13,14,16]. Second, comparing the nearly constant $N_s$ at $E_F$<0.3eV for sp³ impurities in Fig. 4 (both theory and experiment) with gradually decreasing $N_s$ for vacancies indicates that the magnetic moment is sensitive to the shape of the DOS peak: the latter is much broader for vacancies [14] resulting in a more gradual decrease of $N_s$ in (2), in agreement with experiment. Third, an alternative mechanism for quenching of the magnetic moments at high $n$ could be the Kondo effect [17,20]. However, the observed cumulative contribution from localized and itinerant moments and their $T$ dependence do not support this scenario (see Supplementary Note 3). Fourth, the discussed threshold doping is achievable in field-effect devices with high-quality gate dielectrics such as boron nitride. This should allow strong suppression of the rapid spin relaxation, characteristic to the existing graphene devices, so that spin currents can be efficiently switched on and off by gate voltage as required for a spin transistor operation.

**Methods**

Our graphene samples were prepared as described in ref. [8]. In brief, we used ultrasonic cleavage of highly-oriented pyrolytic graphite in an organic solvent, N-methylpyrrolidone (NMP), following the procedure described in ref. [33]. The resulting laminates consist of predominantly 30 to 50 nm crystallites, aligned parallel to each other and electronically decoupled [8]. During filtration of the graphene suspension in NMP, the crystallites do not re-stack into graphite but remain rotationally disordered, with the average separation between graphene flakes ≈0.36 nm, significantly larger than the interplane distance in graphite.

Vacancies were introduced by proton irradiation in a 500kV ion implanter, at room $T$, and current densities <0.2 µA/cm$^2$. The energy (between 350 to 400 keV), fluence (1.2 to 11 ×10$^{15}$ cm$^{-2}$) and other irradiation parameters for each sample were chosen to achieve a desired defect density and ensure uniformity of defect distribution [8]. Proton irradiation under such conditions produces predominantly single vacancies [12]; their number for each sample was estimated on the basis of the corresponding fluence, surface area and the number of defects created by each proton as obtained from computer simulations (SRIM software package). Further details can be found in ref. [8]. We note that impacts of energetic protons on graphene may also give rise to a finite concentration of other defects, such as divacancies and carbon adatoms, but their effect on magnetism is negligible, as the former do not possess magnetic moment and the latter are mobile at room temperature and tend to cluster with the loss of magnetism [34].

In our SQUID measurements, the field was usually applied parallel to the lamination direction to avoid diamagnetism that is strong in the perpendicular field [8]. A small remnant diamagnetic background due to some misalignment was linear with $B$ up to 7T at all $T$ and subtracted for clarity from the plotted data.

To monitor the changes in $n$ induced by doping we used Hall measurements and Raman spectroscopy (Supplementary Note 1). The best results in terms of doping control, reproducibility and homogeneity were obtained with HNO$_3$. The effect of NO$_2$ gas on the reported magnetism was qualitatively similar but changes in $n$ were more difficult to control and doping was less uniform. The maximum doping level we could achieve corresponded to $n$≈2·10$^{13}$ cm$^{-2}$ or $|E_F| = (\pi n)^{1/2} \hbar v_F \approx 0.54$ eV. Importantly, once exposed to HNO$_3$, the carrier concentration remained stable for many days at room $T$, in agreement with relatively high adsorption energy for this molecule [23] but recovered the initial value under mild annealing in argon. Further details of the doping procedure and analysis of carrier concentrations can be found in Supplementary Information.

**Acknowledgements.** This work was supported by the UK Engineering and Physical Sciences Research Council. A.H.C.N. acknowledges support from NRF-CRP (R-144-000-295-281). A.V.K., O.L. and J.K. acknowledge support from the Academy of Finland through projects 218545 and 263416.

# Supplementary Information

**#1. Molecular Doping and Determination of Carrier Concentrations.** Molecular doping has been shown to be an effective way of changing carrier concentration in graphene, with water, $NO_2$ gas and nitric acid ($HNO_3$) identified as acceptors (hole dopants) and ammonia, anisole, aniline and N,N,N',N'-tetramethyl-p-phenylenediamine (TMPD) as donors (electron dopants) [23, 25,26, 35-39]. After testing the above molecules on graphene laminates, nitric acid was found to be the only molecule providing sufficiently high carrier concentrations, $n$ (up to $n \approx 2 \cdot 10^{13}$ cm$^{-2}$, see below), excellent stability ($n$ remained same for several weeks) and uniform distribution (as manifested in same $n$ measured in different locations on graphene laminates). To achieve gradual changes in $n$, the laminates were exposed to $HNO_3$ diluted with water at different molar concentrations of 0.1 to 15M.

Gaseous $NO_2$ produced qualitatively similar results (in terms of both measured $n$ and the effect on magnetisation) but the doping level was difficult to control and it showed significant non-uniformity. Ammonia did not produce any measurable permanent changes in $n$, presumably because of the difficulty in diffusing ammonia molecules between graphene layers and also much smaller adsorption energy compared to $NO_2$ [38]. The only donor that produced systematic changes in $n$ was aniline [36] but the maximum possible $n$ was small (~ $5 \times 10^{12}$ cm$^{-2}$) and resulted in only a small reduction in paramagnetic magnetisation, similar to the effect of hole doping at such $n$. Another known electron dopant N,N,N',N'-tetramethyl-p-phenylenediamine (TMPD) [39] was found to exhibit weak paramagnetism of its own and, therefore, was avoided.

To monitor the changes in $n$ we used Hall effect measurements. They were done on 3×3 mm samples cut out of a 2 cm-diameter laminate obtained after filtration. The concentration was calculated as $n = I \cdot B/(V_H \cdot e \cdot N_l)$, where $I$ is the applied current, $B$ the magnetic field, $V_H$ the Hall voltage and $N_l$ the number of graphene layers in the laminate. The latter was calculated from the measured sample thickness, typically 3-5μm, and the average separation between the layers $d \approx 0.36$ nm, as determined using X-ray diffraction.

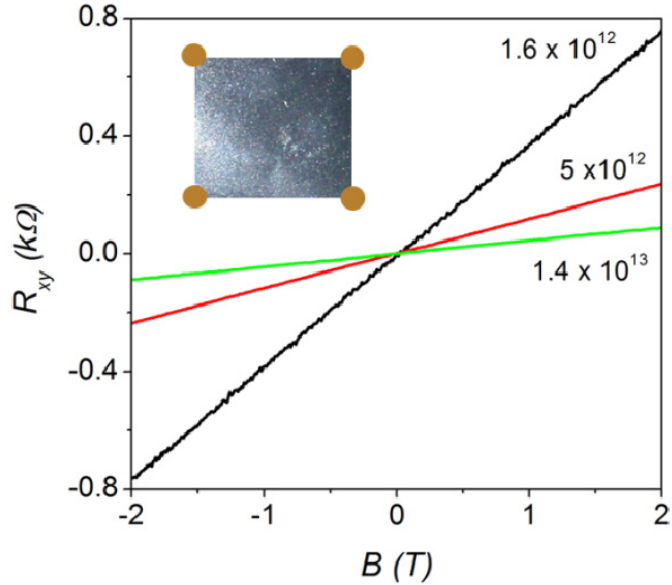

**Supplementary Figure S1. Carrier concentrations in doped graphene laminates.** Hall resistance per graphene layer in the laminate ($R_{xy}$) as a function of applied magnetic field for $HNO_3$ doped graphene. Labels show corresponding carrier concentrations in cm$^{-2}$. *Inset*: Hall measurements were carried out in the van-der-Pauw geometry.

In addition to the Hall measurements, we used Raman spectroscopy that is also sensitive to $n$ [25, 40, 41]. Raman measurements were carried out using a Renishaw spectrometer with a green laser. As graphene crystallites in the laminates are small (typically 30 to 50 nm), they produce a significant broadening of all Raman peaks, which makes it difficult to detect exact shifts in their positions that would otherwise yield $n$. Therefore we used graphene mechanically exfoliated onto Si/SiO$_2$ and treated these samples in parallel with graphene laminates used for the Hall and magnetisation measurements. Raman spectroscopy on such isolated samples was found to be sufficiently sensitive to detect changes in $n$ even for small changes in $HNO_3$ concentration.

Examples of Raman spectra after exposure to different concentrations of $HNO_3$ are shown in Supplementary Figure S2. We examined the shift and intensity variations of the G peak that is most sensitive to $n$ [25,40,41]. The changes were translated into $n$ by using the calibration data for gated graphene [41]. The absence of the D peak either before or after exposure to $HNO_3$ (not shown) confirmed that $HNO_3$ molecules were physisorbed on graphene and not chemisorbed, as the D peak is known to be a measure of short-range defects such as sp$^3$ bonds [42]. Corresponding Fermi energies for each $n$ were calculated using the relation $E_F = (\pi n)^{1/2} \hbar v_F$, where $v_F = 1.1 \cdot 10^6$ ms$^{-1}$ is the Fermi velocity. We emphasize that the results, shown in Supplementary Figure S2, are in good agreement with the concentrations obtained from our Hall measurements for the same exposure to $HNO_3$.

As a further check, we also measured Raman spectra directly on graphene laminates at high doping (HNO$_3$ concentrations of 11-15M), where the shift in the G peak position was sufficiently large (peak at 1600-1605 cm$^{-1}$) to overcome the broadening effect. Again, the results were consistent with the Hall measurements and with the values of $n$ obtained from the Raman spectra on exfoliated monolayers. For these particular measurements, the extracted values of $n$ were in a range of (1-2)·10$^{13}$ cm$^{-2}$.

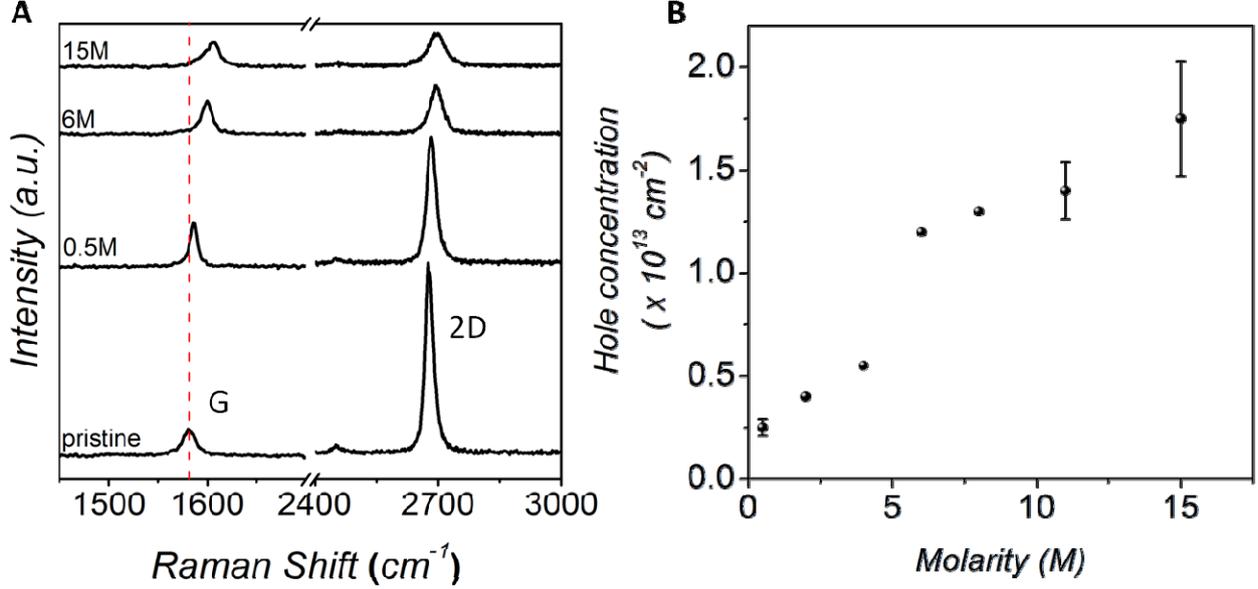

**Supplementary Figure S2. Evolution of Raman spectra with doping.** A – Examples of Raman spectra of monolayer graphene mechanically exfoliated onto Si/ SiO$_2$ substrate and exposed to nitric acid at different concentrations; labels show molar concentrations of HNO$_3$. Corresponding G-peak positions are 1581±0.5, 1586±0.5, 1599±0.5 and 1603±1 cm$^{-1}$. B – Carrier concentrations found from the G peak positions.

#2. Density of states for graphene with sp$^3$ impurities

The simplest description of the electronic structure of adsorbates covalently bound to carbon atoms in graphene is based on the hybridization model, where two types of electronic states are postulated: localized (responsible for the formation of mid-gap states) and itinerant, or conduction electron states [15,31,43,44]. For the hybridization model the Green's function of the localized states can be expressed as

$$G_d(E) = \frac{1}{E - E_d - \sum_k \frac{|V_k|^2}{E - t(k)}}$$

where $t(k)$ is the band-electron dispersion, $V_k$ is the momentum-dependent hybridization parameter (see Ref. [43], Eq.(17)). Assuming for simplicity $V_k = V = const$ we find

$$G_d(E) = \frac{1}{E - E_d - V^2 g_0(E)}$$

where $g_0(E)$ is the on-site Green's function of conduction electrons for an ideal crystal (see Ref. [43], Eq.(6)). For the latter we have chosen a simplified model of the electron energy spectrum with the linear-in-energy density of states cut at the half-bandwidth ("graphene" model of Ref. [43]). Thus, we have:

$$N(E) = -\frac{1}{\pi}\operatorname{Im} G_d(E) = \frac{a|E|}{\left[E\left(1 - a\ln\left|\frac{E^2}{1-E^2}\right|\right) + E_d\right]^2 + (\pi a E)^2}$$

where energies $E, E_d$ are expressed in terms of the half-bandwidth, $D = 3t$, $a = (V/D)^2$.

**# 3. Temperature dependence of magnetic susceptibility and remarks on Kondo effect**

The central result of our study is the observation of quenching of defects' magnetic moments at sufficiently high doping, where the Fermi level is shifted to $E_F \geq 0.45$ eV. We attribute this effect to double occupation (with both up and down spins) of the localised states created by the defects (see main text). An alternative explanation of this effect could be the much-discussed Kondo effect in graphene [17,20,45-47]. The Kondo effect in irradiated graphene was claimed by Chen et al [45] on the basis of the observed logarithmic corrections to the resistivity of ion-irradiated graphene samples. However, the same experimental data allow an alternative explanation based on weak localization corrections [46,47]. Also, the *s-d* exchange coupling between localized and conduction electron spins has been shown to be *ferromagnetic*, i.e. the alignment of the localized and conduction electron spins is expected to be parallel [17], which rules out the Kondo effect.

Irrespective of this theoretical work, our experimental observation that the spins of localized (σ-band) and itinerant (π-band) electrons do not cancel each other shows that they are, indeed, *parallel*. In addition, the Kondo effect − if relevant - would show up in magnetisation measurements as deviations of the $T$ dependence of the magnetic susceptibility χ from the Curie law $\chi \propto 1/T$, at $T$ below the Kondo temperature [48]. Therefore, to distinguish between quenching of the magnetic moments due to doping and due to possible Kondo effect, we measured $\chi(T) = M(T)/H$, for both types of studied defects (vacancies and $sp^3$ impurities). Here $M(T)$ is the measured magnetic moment and $H$ the applied field. Supplementary Figure S3 shows the results for one of the irradiated samples and Supplementary Figure S4 for a sample with $sp^3$ defects. In both cases the Curie law holds accurately down to 2K, the lowest $T$ in our experiments, without any noticeable offset from zero. This also contradicts the existence of antiferromagnetic coupling (corresponding to the Kondo effect). Therefore, our data clearly do not support the Kondo scenario.

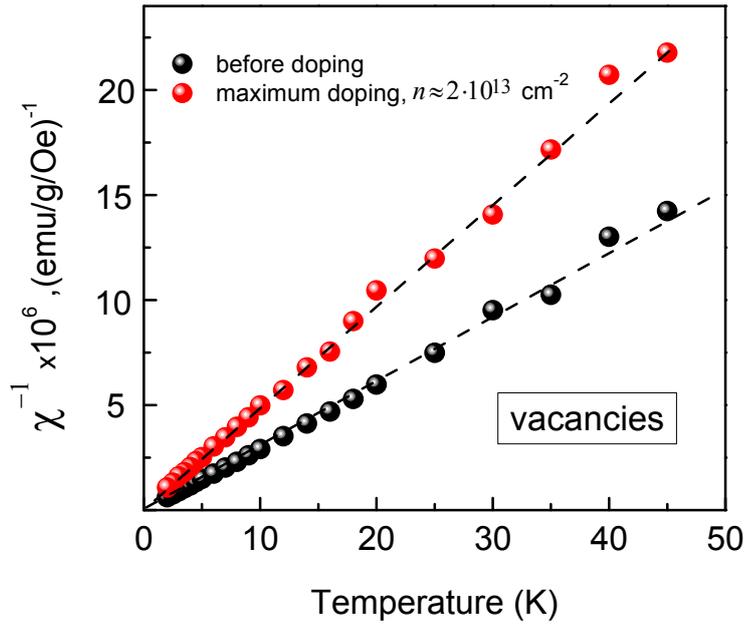

**Supplementary Figure S3. Absence of deviations from the Curie law for graphene with vacancies.** Inverse paramagnetic susceptibility for an irradiated sample also used in the main text. Measurements were done in parallel field $H$= 3 kOe. Constant diamagnetic background subtracted.

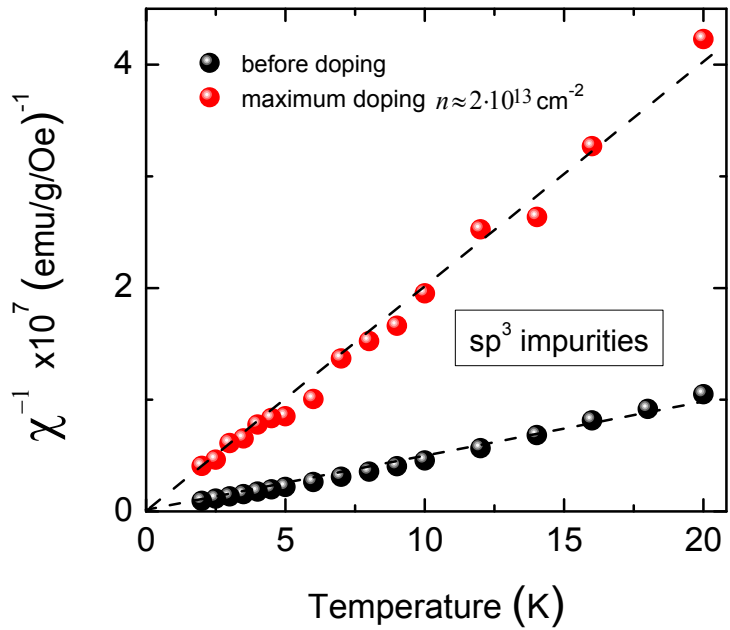

**Supplementary Figure S4. Absence of deviations from the Curie law for graphene with sp³ impurities.** This sample was heat-treated at 350°C (see main text). $H$= 5 kOe.

**Supplementary References**